\documentclass[aps,prb,twocolumn,superscriptaddress,groupedaddress]{revtex4-2}  
\usepackage{amsmath}
\usepackage{graphicx}
\usepackage{subfigure}
\usepackage{color}
\usepackage{amsfonts}
\usepackage{tikz}
\usepackage{esint}
\usepackage{mathrsfs}
\usepackage{hyperref}
\hypersetup{
     colorlinks = true,
     citecolor  = blue,  
     linkcolor  = blue, 
     urlcolor = blue,
     filecolor = blue
}
\begin{document}

\title{Gutzwiller approximation approach to the SU(4) $t$-$J$ model}

\author{Jia-Cheng He}
\email{jche14@fudan.edu.cn}

\author{Jie Hou}

\author{Yan Chen}
\email{yanchen99@fudan.edu.cn}
\affiliation{Department of Physics and State Key Laboratory of Surface Physics, Fudan University, Shanghai 200433, China}
\date{\today}

\begin{abstract}
We develop the Gutzwiller approximation method to obtain the renormalized Hamiltonian of the SU(4) $t$-$J$ model with the corresponding renormalization factors. Subsequently, a
mean-field theory is employed on the renormalized Hamiltonian of the model on the honeycomb lattice under the scenario of a cooperative condensation of carriers moving in the
resonating valence bond state of flavors. In particular, we find that the extended $s$-wave superconducting state is more favorable than the $d\pm id$-wave superconducting state in the doping range close to quarter filling. The pairing states of the SU(4) case reveal the property that the spin-singlet pairing and the spin-triplet pairing can coexist simultaneously. Our results might provide new insights into the twisted bilayer graphene system.
\end{abstract}

\maketitle

\section{Introduction}
When the resonating valence bond (RVB) state or quantum spin liquid (QSL) is doped sufficiently, the preexisting magnetic singlet pairs of the RVB state become charged
superconducting pairs. This idea is proposed for the mechanism of superconductivity of the Copper oxide superconductors \cite{P.W.Anderson1987}. Moreover, it is a usual way to realize the QSL by imposing the frustrated lattices on the SU(2) spin system. Despite many studies on this idea, it is still challenging to find the real material candidates for the QSL.

To realize the QSL, it is an alternative approach that the spin symmetry group of the electrons is extended from SU(2) to SU(N). For the sake of more substantial quantum fluctuations, we
expect the QSL in SU(N) ``spin" system with large N (N$>$2) even on unfrustrated bipartite lattices, like the honeycomb lattice. The SU(2) Heisenberg model can be generalized to the SU(N)
one with the consideration of orbital degeneracy in addition to spin degeneracy \cite{F.C.Zhang1998PRLSU4, ZhichaoZhouPRB2016, CenkeXu}. The QSL realized through orbital degeneracy can be
named quantum spin-orbital liquid. With the combination of the octahedral ligand field and strong spin-orbital coupling, the SU(4) Heisenberg model on the honeycomb lattice emerges in
$\alpha$-ZrCl$_{3}$ at quarter filling \cite{GeorgeJackeli2018PRL}.
We can naturally obtain the corresponding SU(4) $t$-$J$ model by slightly doping holes on the SU(4) Heisenberg model. It is interesting whether there hosts superconducting state in the SU(4)
$t$-$J$ model, just like the superconductivity mechanism related to the QSL.

Recently, the discovery of the superconductivity and correlated insulating state in the twisted bilayer graphene (TBG) \cite{Y.CaoSu,Y.CaoCo} has driven a surge of theoretical
works on this field \cite{CenkeXu,LiangFuPRB2018, LiangFuPRX2018,
D.N.Sheng2019,T.Senthilfirstpaper2018,PatrickA.Lee2018,S.A.Kivelson2018,FanYangPRL2018,PhysRevX.8.031088,PhysRevLett.122.026801,PhysRevB.98.241407,PhysRevB.99.094521,PhysRevB.99.121407,PhysRevB.99.144507}.
The insulating state and superconductivity are closely related to the flat bands. A sufficiently large energy gap separates these flat bands well from the excited bands. The phenomena in
TBG are very like that of cuprates in many aspects. The correlated insulating state appears at the filling of $\pm 2e$ per superlattice unit cell. After doping, the two superconducting
domes are observed on both sides of the insulating state. Moreover, $T_{c}/T_{F}$, the ratio of the critical superconducting temperature $T_{c}$ to the Fermi temperature $T_{F}$ in TBG
is even higher than that of cuprates, indicating very strong electron-electron interactions \cite{Y.CaoSu,MatthewYankowitz2019}. The insulating and superconducting phases share similar
energy scales \cite{MatthewYankowitz2019}, and this constrains models in which the superconductivity arises as a daughter-state of the insulator. Another important point of TBG is the
twofold valley degeneracy, where the intervalley scattering requires a large momentum transfer compared to the mini Brillouin zone. The local density states of the flat bands in
TBG are highly concentrated in the AA stacking regions, which form a triangular lattice \cite{Y.CaoCo}. However, according to the symmetry representations, all the triangular lattice
models for TBG are ruled out \cite{T.Senthilfirstpaper2018}. Any tight-binding model proposed for TBG must correspond to Wannier orbitals forming a honeycomb lattice
\cite{T.Senthilfirstpaper2018,LiangFuPRB2018}. These orbitals centered at AB/BA regions have nontrivial shapes: their weight is mainly concentered in the AA stacking regions
\cite{LiangFuPRB2018,LiangFuPRX2018,T.Senthilfirstpaper2018,PhysRevX.8.031088}. A two-orbital extended Hubbard model on the honeycomb lattice may be a good starting point to study the
correlated states in TBG \cite{LiangFuPRB2018,T.Senthilfirstpaper2018}. Here the honeycomb lattice model can accommodate up to eight electrons per unit cell, corresponding to the complete
filling of miniband in TBG. The charge neutrality point of TBG corresponds to four electrons per unit cell in this model \cite{LiangFuPRB2018}. It is more appropriate to study the SU(4) Hubbard model or SU(4) $t$-$J$ model on the honeycomb lattice, just like the Refs. \cite{D.N.Sheng2019,PhysRevB.98.241407}.

In addition to possible explanations for the real materials, the study of the SU(4) $t$-$J$ model is a natural generalization of the famous SU(2) $t$-$J$ model, which may provide a fascinating insight into unconventional superconductivity. However, similar to the SU(2) $t$-$J$ model, the SU(4) $t$-$J$ model is too difficult to solve analytically. The application of Gutzwiller approximation (GA) to the SU(2) $t$-$J$ model is quite successful \cite{PatrickA.LeeRMP2006}. The renormalization factors of the SU(2) $t$-$J$ model are essential in the solution, but the renormalization factors of the SU(4) case have been unknown due to the complexity of fermions possessing more than two internal components.

In this paper, our starting point is a SU(4) Hubbard model on the honeycomb lattice, with four flavors of single-electron states per site, including both the spin and the orbital. We
only consider the strong correlation limit case, the SU(4) $t$-$J$ model. Moreover, we develop the GA method \cite{F.C.Zhang} to obtain a renormalized Hamiltonian of the SU(4) $t$-$J$
model with the Gutzwiller renormalization factors. Following our idea, one can easily obtain the Gutzwiller renormalization factors of the SU(N) $t$-$J$ model. As shown in Fig.
\ref{FirstPicture}(c), the Gutzwiller renormalization factors of the SU(4) $t$-$J$ model have a new property compared to the SU(2) case. Then we resort to a further mean-field
approximation to analytically deal with this renormalized Hamiltonian under the scenario of the RVB state. Our numerical results found that the ground state in the doping area close to
quarter filling is the extended $s$-wave state, and superconductivity almost disappears in the doping area close to half-filling. The pairing states of the SU(4) case reveal the property
that the spin-singlet pairing and the spin-triplet pairing can exist simultaneously.

This paper is organized as follows. In Section \ref{sec:modelandgfactors} the model is introduced, and the renormalized Hamiltonian is derived. In Section \ref{sec:BdGway} a mean-field
theory is utilized to handle the renormalized Hamiltonian on the honeycomb lattice. Section \ref{sec:numericalresults} presents the numerical results. The discussion and conclusion will be given in Section \ref{sec:discusionandconclusion}.

\section{The model and the renormalization factors}\label{sec:modelandgfactors}
As introduced above, our starting point is a two-orbital Hubbard model on the honeycomb lattice. The SU(4) Hubbard model with four flavors of single-electron states on each site:
\begin{equation}\label{eq: su(4) Hubbard model}
H^{SU(4)}=-t\sum_{\langle i,j\rangle}\sum_{\alpha=1}^{4}(c^{\dagger}_{i,\alpha}c_{j,\alpha}+\mathrm{h.c.})+U\sum_{j}(\sum_{\alpha=1}^{4}\hat{n}_{j,\alpha})^{2},
\end{equation}
where $\langle i,j\rangle$ represents the nearest-neighbor sites pair and $\hat{n}_{j,\alpha}=c^{\dagger}_{j,\alpha}c_{j,\alpha}$. Let's consider the limit of strong correlation of the
SU(4) Hubbard model, i.e., $U\gg t$, with lightly hole doping away from integer filling number $\nu_{0}=$ 1, 2 or 3 (i.e., the number of electrons on each site is $\nu_{0}$) and follow the
procedure of the perturbation theory with the trick of Fierz identity \cite{IgorF.Herbut,LucileSavary,CenkeXu}. Then we obtain the SU(4) $t$-$J$ model \cite{CenkeXu}:
\begin{flalign}\label{eq:su(4) $t$-$J$ model}
H_{t-J}^{SU(4)}=&-\sum_{\langle i,j\rangle}\sum_{\alpha=1}^{4}P_{G}t(c^{\dagger}_{i,\alpha}c_{j,\alpha}+\mathrm{h.c.})P_{G}
                     \nonumber\\
&+J\sum_{\langle i,j\rangle}\sum^{15}_{a=1}\hat{T}^{a}_{i}\hat{T}^{a}_{j},
\end{flalign}
and its wavefunction $|\Psi\rangle=P_{G}|\Psi_{0}\rangle$, where $J=t^{2}/4U$, $\hat{T}_{i}^{a}=\sum_{\alpha\beta}c_{i, \alpha}^{\dagger} \Gamma_{\alpha \beta}^{a} c_{i, \beta}$,
$\Gamma^{a}$ which is the generator of SU(4) group satisfying the relation $\mathrm{Tr}\left(\Gamma^{a}\Gamma^{b}\right)=4\delta_{ab}$, $|\Psi_{0}\rangle$ being unprojected wave function
and $P_{G}$ being the projection operator which excludes the particle occupancy states that stay at the very high energy levels arising from the Hubbard $U$ term in Eq. (\ref{eq: su(4)
Hubbard model}).

Essentially the effect of $P_{G}$ projection operator originates from the configuration minimizing the interaction energy (i.e., $U$ term in Eq. (\ref{eq: su(4) Hubbard model})). We need
to find the wave function configuration minimizing the interaction energy of Eq. (\ref{eq: su(4) Hubbard model}) and let us call this \emph{wave function correlation configuration}. This
wave function correlation configuration is equal to $P_{G}|\Psi_{0}\rangle$. For example, in the case of hole doping away from integer filling number $\nu_{0}=2$, the wave function
correlation configuration indicates that there are no triple occupancy and quadruple occupancy on each site after $P_{G}$ projection, and not so obviously the wave function correlation
configuration tells us the fact that empty occupancy is also excluded after $P_{G}$ projection. Therefore, only single occupancy and double occupancy are allowed after $P_{G}$
projection in the case of hole doping away from integer filling number $\nu_{0}=2$. Following the same steps, we can obtain the result that in the case of lightly hole doping away from integer filling number $\nu_{0}$ ($\nu_{0}=$ 1, 2, or 3), only $\nu_{0}$ occupancy and $\nu_{0}-1$ occupancy are permitted in the wave function correlation configuration.

The GA \cite{F.C.Zhang, B.Edegger} is a method that the correlation effect is absorbed in the renormalized factors. Here we only consider the on-site correlation effect. The GA uses the
expectation value of the operator $\hat{O}$ within the unprojected state $|\Psi_{0}\rangle$ multiplying a statistical weight $g_{O}$ defined by
$g_{O}\approx\langle\hat{O}\rangle_{\Psi}/\langle\hat{O}\rangle_{\Psi_{0}}$ to approximate the expectation value of the operator $\hat{O}$ within the projected state
$|\Psi\rangle=P_{G}|\Psi_{0}\rangle$. Here $\langle\cdot\cdot\cdot\rangle_{\Psi}$ ($\langle\cdot\cdot\cdot\rangle_{\Psi_{0}}$) represents the expectation value with respect to the wave
function $|\Psi\rangle$ ($|\Psi_{0}\rangle$). The starting point of GA is the calculation of the probabilities for the occupancy at any site $i$. In this paper, for simplicity, we only
consider the situation of a homogeneous wave function with fixed particle number and flavor symmetry. We need to calculate the probabilities of a site occupied by any number of electrons
with different flavors. For the case of hole doping away from integer filling number $\nu_{0}$, the probabilities for different occupancies on site $i$ in $|\Psi\rangle$ and $|\Psi_{0}\rangle$
are listed in Table \ref{table:occupancyprobability}. We should consider five kinds of occupancies: empty occupancy, single occupancy, double occupancy, triple occupancy and quadruple
occupancy as listed in the first column of Table \ref{table:occupancyprobability}. We use the notation $n^{0}_{i\alpha}$ to represent the density before projection as listed in the third
column and here we let $n^{0}_{i\alpha}=n/4$ and $n=\langle\hat{n}_{i}\rangle_{\Psi}=4\langle\hat{n}_{i\alpha}\rangle_{\Psi}$ with $\hat{n}_{i}=\sum_{\alpha}\hat{n}_{i\alpha}$.

\begin{table*}
\caption{\label{table:occupancyprobability}Probabilities of different occupancies for the case of hole doping away from integer filling number $\nu_{0}$ ($\nu_{0}=$ 1, 2 or 3) for the SU(4) $t$-$J$ model. $n^{0}_{i\alpha}$ represents the density before projection and here we let $n^{0}_{i\alpha}=n/4$ and
$n=\langle\hat{n}_{i}\rangle_{\Psi}=4\langle\hat{n}_{i\alpha}\rangle_{\Psi}$ with $\hat{n}_{i}=\sum_{\alpha}\hat{n}_{i\alpha}$.}
\begin{ruledtabular}
\begin{tabular}{lll}
Occupancy on site $i$ &Probabilities in $|\Psi\rangle$ &Probabilities in $|\Psi_{0}\rangle$ \\
\hline
\\
$\langle\prod^{\nu_{0}}_{x=1}\hat{n}_{ix}\prod^{4}_{y=\nu_{0}+1}(1-\hat{n}_{iy})\rangle$ & $\frac{1-\nu_{0}+n}{C_{4}^{\nu_{0}}}$ &
$\prod^{\nu_{0}}_{x=1}n_{ix}^{0}\prod^{4}_{y=\nu_{0}+1}(1-n_{iy}^{0})$ \\
\\
$\langle\prod^{\nu_{0}-1}_{x=1}\hat{n}_{ix}\prod^{4}_{y=\nu_{0}}(1-\hat{n}_{iy})\rangle$ & $\frac{\nu_{0}-n}{C_{4}^{\nu_{0}-1}}$ &
$\prod^{\nu_{0}-1}_{x=1}n_{ix}^{0}\prod^{4}_{y=\nu_{0}}(1-n_{iy}^{0})$ \\
\\
$\langle\prod^{\rho\ne\nu_{0},\nu_{0}-1}_{x=1}\hat{n}_{ix}\prod^{4}_{y=\rho+1}(1-\hat{n}_{iy})\rangle$ & $0$ &
$\prod^{\rho\ne\nu_{0},\nu_{0}-1}_{x=1}n_{ix}^{0}\prod^{4}_{y=\rho+1}(1-n_{iy}^{0})$ \\
\end{tabular}
\end{ruledtabular}
\end{table*}

\begin{figure}
\centering
\includegraphics[width=0.48\textwidth]{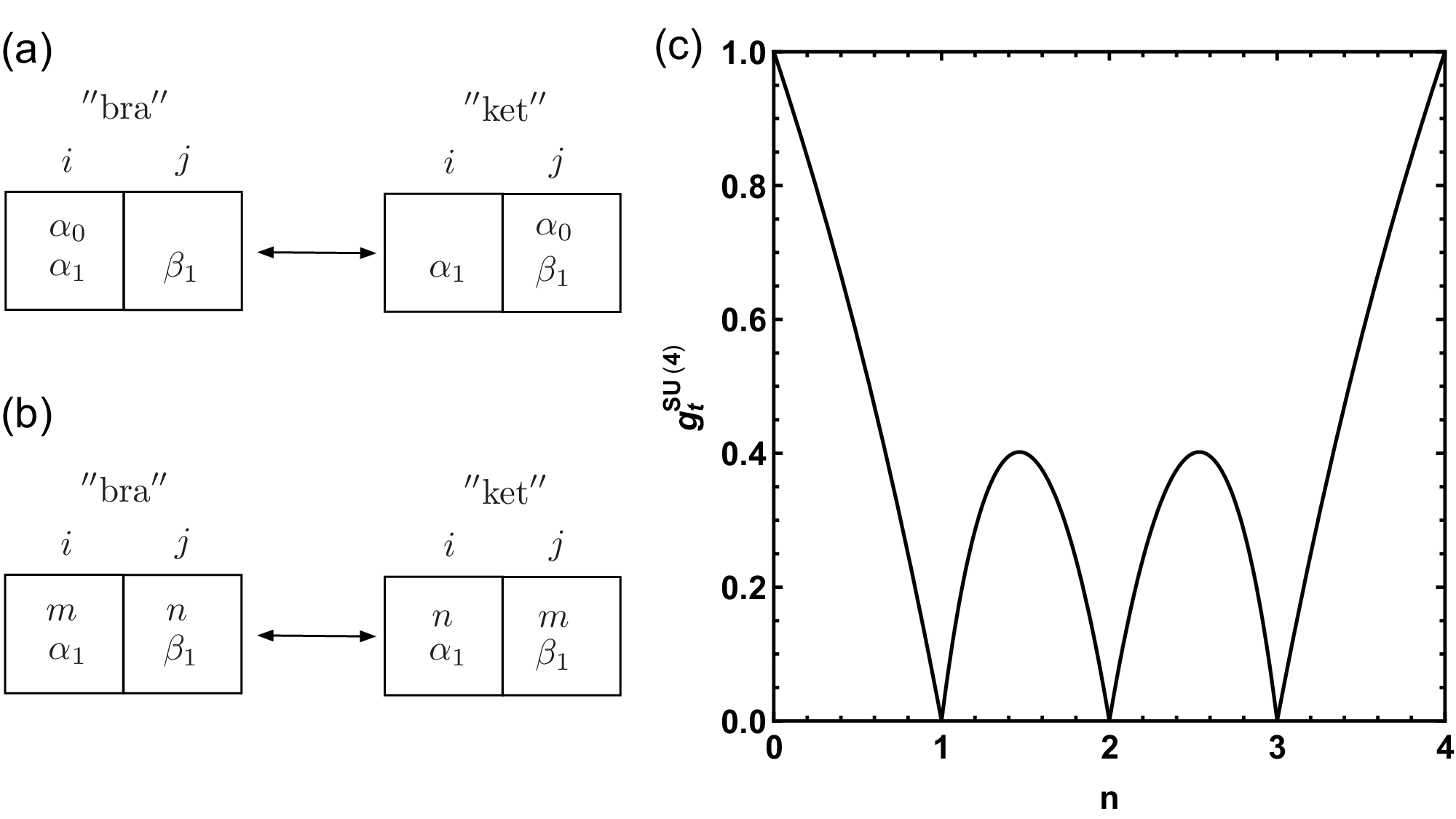}
\caption{\label{FirstPicture} (a) The permitted hopping progress in the case of hole doping away from integer filling number $\nu_{0}=2$, where $\alpha_{0}$ is different from both $\alpha_{1}$
and $\beta_{1}$. (b) The permitted flavor exchange progress in the case of hole doping away from integer filling number $\nu_{0}=2$, where $\alpha_{1}$ ($\beta_{1}$) is different from both $m$
and $n$. (c) $g_{t}^{SU(4)}(n,\nu_{0})$ as the function of electron density $n$ is symmetric about the point $n=2$.}
\end{figure}

Using Table \ref{table:occupancyprobability} and following the procedure of GA, we can obtain the renormalization factors related to the statistical weight ratios and remember that the
ratio $g_{O}$ should be calculated by considering the probability amplitudes of ``bra" and ``ket" configurations that contribute \cite{B.Edegger}. For the case of hole doping away from integer
filling number $\nu_{0}$, following the procedure of GA we can obtain the relation about the hoping term $\langle c^{\dagger}_{i,\alpha}c_{j,\alpha}\rangle_{\Psi}\approx
g_{t}^{SU(4)}(n,\nu_{0})\langle c^{\dagger}_{i,\alpha}c_{j,\alpha}\rangle_{\Psi_{0}}$, where
\begin{flalign}\label{SU4gtfactor}
g_{t}^{SU(4)}(n,\nu_{0})=\frac{\nu_{0}(5-\nu_{0})[ n-(\nu_{0}-1)](\nu_{0}-n)}{n(4-n)},
\end{flalign}
is the renormalization factor for hopping terms. The corresponding hopping progress is described by Fig. \ref{FirstPicture}(a) as $\nu_{0}=2$ case. In Fig. \ref{FirstPicture}(a)
$\alpha_{0}$, $\alpha_{1}$ and $\beta_{1}$ represent the flavors of electrons, where $\alpha_{0}$ is different from both $\alpha_{1}$ and $\beta_{1}$. Then, we consider the interaction
term $\sum^{15}_{a=1}\hat{T}^{a}_{i}\hat{T}^{a}_{j}$ which can be rewritten as
$\sum^{15}_{a=1}\hat{T}^{a}_{i}\hat{T}^{a}_{j}=\frac{1}{2}\sum_{m<n}\left(\hat{\mathcal{T}}_{i}^{mn+}\hat{\mathcal{T}}_{j}^{mn-}+\hat{\mathcal{T}}_{i}^{mn-}\hat{\mathcal{T}}_{j}^{mn+}\right)
+\sum_{l=1}^{3}\hat{\mathcal{T}}_{i}^{l}\hat{\mathcal{T}}_{j}^{l}$
where $\hat{\mathcal{T}}_{i}^{mn+}=\sqrt{8}c_{im}^{\dagger}c_{in}$, $\hat{\mathcal{T}}_{i}^{mn-}=\sqrt{8}c_{in}^{\dagger}c_{im}$ and
$\hat{\mathcal{T}}_{i}^{l}=\sqrt{8}\sum_{\alpha\beta}c_{i\alpha}^{\dagger}T_{\alpha\beta}^{l+1}c_{i\beta}$,  $T^{2}=\mathrm{diag}\{1,-1,0,0\}/\sqrt{4}$,
$T^{3}=\mathrm{diag}\{1,1,-2,0\}/\sqrt{12}$, $T^{4}=\mathrm{diag}\{1,1,1,-3\}/\sqrt{24}$. We should keep in mind that when we calculate the renormalized factors about the interaction
term we suppose lightly hole doping away from integer filling number $\nu_{0}$, because this is the precondition from Hubbard model to obtain the $J$ term. Therefore we obtain
$\langle\hat{\mathcal{T}}_{i}^{mn+}\hat{\mathcal{T}}_{j}^{mn-}\rangle_{\Psi}\approx
g_{S}^{SU(4)}(n,\nu_{0})\langle\hat{\mathcal{T}}_{i}^{mn+}\hat{\mathcal{T}}_{j}^{mn-}\rangle_{\Psi_{0}}$, where
\begin{flalign}\label{SU4gsfactor}
g_{S}^{SU(4)}(n,\nu_{0})=\left[\frac{1-\nu_{0}+n}{C_{4}^{\nu_{0}}\left(\frac{n}{4}\right)^{\nu_{0}}\left(1-\frac{n}{4}\right)^{4-\nu_{0}}}\right]^{2}.
\end{flalign}
For $\langle\hat{\mathcal{T}}_{i}^{mn+}\hat{\mathcal{T}}_{j}^{mn-}\rangle_{\Psi}$ term the corresponding flavor exchange progress is described by Fig. \ref{FirstPicture}(b) as
$\nu_{0}=2$ case. In Fig. \ref{FirstPicture}(b) $m$, $n$, $\alpha_{1}$ and $\beta_{1}$ represent the flavors of electrons, where $\alpha_{1}$ ($\beta_{1}$) is different from both $m$ and
$n$. And one can prove that
$\langle\hat{\mathcal{T}}_{i}^{l}\hat{\mathcal{T}}_{j}^{l}\rangle_{\Psi}=g_{S}^{SU(4)}(n,\nu_{0})\langle\hat{\mathcal{T}}_{i}^{l}\hat{\mathcal{T}}_{j}^{l}\rangle_{\Psi_{0}}$. Namely,
$g_{S}^{SU(4)}(n,\nu_{0})$ is the renormalization factor for the interaction terms. $g_{t}^{SU(4)}(n,\nu_{0})$ as the function of electron density $n$ is shown in Fig.
\ref{FirstPicture}(c). Since the SU(4) Hubbard model described by Eq. (\ref{eq: su(4) Hubbard model}) has particle-hole symmetry, both $g_{t}^{SU(4)}(n,\nu_{0})$ and
$g_{S}^{SU(4)}(n,\nu_{0})$ are symmetric about the point $n=2$. In Fig. \ref{FirstPicture}(c), $g_{t}^{SU(4)}$ is zero at integer filling numbers $\nu_{0}=1,2,3$ corresponding to the
Mott insulating states. It enables the renormalization factor $g_{t}^{SU(4)}$ nonmonotonic that the emergent flavors fluctuation background arises in the cases of $\nu_{0}=2$ and
$\nu_{0}=3$. The underlying physics of the nonmonotonic property is that the flavors background plays both the role increasing the resource of mobile electrons and the role blocking the
movement of mobile electrons. Thus we obtain the renormalized Hamiltonian
\begin{flalign}\label{eq:g su(4) $t$-$J$ model}
H^{\prime}=&-tg_{t}^{SU(4)}H_{t}+Jg_{S}^{SU(4)}H_{S},
\end{flalign}
where $H_{t}=\sum_{\langle i,j\rangle}\sum_{\alpha=1}^{4}(c^{\dagger}_{i,\alpha}c_{j,\alpha}+ \mathrm{h.c.})$ and $H_{S}=\sum_{\langle
i,j\rangle}\sum^{15}_{a=1}\hat{T}^{a}_{i}\hat{T}^{a}_{j}$.

\section{The Bogoliubov-de Gennes method}\label{sec:BdGway}
We propose that the RVB state is the ground state of the SU(4) $t$-$J$ model on the honeycomb lattice near the integer filling number. Before proceeding further, let us examine the interaction term of the SU(4) $t$-$J$ model. With the help of Fierz identity the interaction term can be rewritten as \cite{CenkeXu}
\begin{equation}\label{eq:pairingform}
\sum_{a=1}^{15} \hat{T}_{i}^{a} \hat{T}_{j}^{a}=-\frac{5}{4}\left(\overrightarrow{\Delta}_{i j}\right)^{\dagger} \cdot \overrightarrow{\Delta}_{i j}+\frac{3}{4}\left(\Delta_{i
j}^{-}\right)^{\dagger} \cdot \Delta_{i j}^{-},
\end{equation}
where $\overrightarrow{\Delta}_{i j}$ and $\Delta_{i j}^{-}$ are the pairing fields with 6 and 10 components, respectively. $\overrightarrow{\Delta}_{i j}$ is the even pairing field
satisfying the relation $\overrightarrow{\Delta}_{i j}=\overrightarrow{\Delta}_{j i}$ and $\Delta_{i j}^{-}$ is the odd pairing field satisfying the relation $\Delta_{i j}^{-}=-\Delta_{j i}^{-}$.
$\overrightarrow{\Delta}_{i j}$ and $\Delta_{i j}^{-}$ have $SO(6)$ rotation symmetry and $SO(10)$ rotation symmetry, respectively. Using $v_{1}$ and $v_{2}$ as the orbital index and
choosing a representation we can obtain $\overrightarrow{\Delta}_{i j}=\tilde{c}_{i}^{t}\left(-\sigma^{32}, i \sigma^{02}, \sigma^{12}, -\sigma^{23}, i\sigma^{20}, \sigma^{21}\right)
\tilde{c}_{j}$ and
 $\Delta_{i j}^{-}=\tilde{c}_{i}^{t}\left(-\sigma^{22}, \vec{\vec{\sigma}}_{\tau}\otimes\vec{\vec{\sigma}}_{\tau}\right) \tilde{c}_{j}$, where $\tilde{c}_{j}=(c_{j\uparrow
 v_{1}},c_{j\uparrow v_{2}},c_{j\downarrow v_{1}},c_{j\downarrow v_{2}})^{t}$, $\sigma^{ab}=\sigma^{a}\otimes\sigma^{b}$, $\sigma^{0}=\mathbf{1}_{2\times2}$ and
 $\vec{\vec{\sigma}}_{\tau}\otimes\vec{\vec{\sigma}}_{\tau}$ being dyadic tensor with $\vec{\vec{\sigma}}_{\tau}=i\hat{\sigma}\sigma^{2}$ and
 $\hat{\sigma}=\sigma^{1}\hat{e}_{1}+\sigma^{2}\hat{e}_{2}+\sigma^{3}\hat{e}_{3}$. About $\overrightarrow{\Delta}_{i j}$ the previous three components describe spin-triplet with
 orbital-singlet pairing field and the last three components describe spin-singlet with orbital-triplet pairing field. For $\Delta_{i j}^{-}$ the first component describes spin-singlet
 with orbital-singlet pairing field and the last nine components describe spin-triplet with orbital-triplet pairing field.

 With the consideration of pairing order parameters
 $\Delta_{\langle i,j\rangle,\alpha\beta}=\langle\Psi_{0}| c_{i, \alpha} c_{j, \beta}|\Psi_{0}\rangle$ we take the expectation value of Eq. (\ref{eq:pairingform}) and obtain
 $-\frac{5}{4}\left(\langle\overrightarrow{\Delta}_{i j}\rangle_{0}\right)^{\dagger} \cdot \langle\overrightarrow{\Delta}_{i j}\rangle_{0}$,
where $\langle\overrightarrow{\Delta}_{i j}\rangle_{0}=\langle\Psi_{0}|\overrightarrow{\Delta}_{i j}|\Psi_{0}\rangle$. Here we neglect the contribution from $\Delta_{i j}^{-}$ due to
$J>0$. $\langle\overrightarrow{\Delta}_{i j}\rangle_{0}$ describes the \emph{smearing} of the pseudo-Fermi surface but is not the superconducting order parameter \cite{F.C.Zhang}.
Following the procedure of GA, for the nearest-neighbor sites $i$ and $j$, the quantity related to the superconducting order parameter is given by
\begin{flalign}
\langle c_{i, \alpha} c_{j, \beta}\rangle_{\Psi}=g_{t}^{SU(4)}(n,\nu_{0})\langle c_{i, \alpha} c_{j, \beta}\rangle_{\Psi_{0}}.
\end{flalign}
With the help of Fierz identity the interaction term can also be rewritten as
\begin{flalign}\label{SU(4) kinetic energy term}
\sum_{a=1}^{15} \hat{T}_{i}^{a} \hat{T}_{j}^{a}=&-\frac{15}{4} \sum_{\alpha} c_{j \alpha}^{\dagger} c_{i \alpha} \sum_{\beta} c_{i \beta}^{\dagger} c_{j \beta}
                       \nonumber\\
&+\frac{1}{4} \sum_{a=1}^{15} \sum_{\alpha \beta} c_{j \alpha} ^{\dagger}\Gamma_{\alpha \beta}^{a} c_{i \beta} \sum_{\delta \gamma} c_{i \delta}^{\dagger} \Gamma_{\delta \gamma}^{a}
c_{j\gamma}.
\end{flalign}
Here we also consider the kinetic energy order parameters $\chi_{\langle i,j\rangle \alpha}=\langle\Psi_{0}|c_{i \alpha}^{\dagger} c_{j \alpha}|\Psi_{0}\rangle$
\cite{F.C.Zhang,Kai-YuYang} and take the expectation value of Eq. (\ref{SU(4) kinetic energy term}) to obtain $-\frac{15}{4}\sum_{\alpha}\chi_{\langle i,j\rangle
\alpha}^{*}\sum_{\beta}\chi_{\langle i,j\rangle\beta}$.
The expectation value of the second term in Eq. (\ref{SU(4) kinetic energy term}) is zero due to the flavor symmetry in our theory. Since the RVB state allows the singlet pairs to move
\cite{P.W.Anderson1987} (the flavor singlet pairs resemble the spin-singlet pairs), the kinetic energy order parameters describe the kinetic energy of the RVB state. For simplicity, in this
paper, we only consider real $\chi_{\langle i,j\rangle ,\alpha}$.

On the honeycomb lattice, we denote by $d$ and $c$ the electron annihilation operators related to the A and B
sublattices of the honeycomb lattice, respectively. We redefine the pairing order parameter:
\begin{equation}
\tilde{\Delta}_{\alpha\beta}^{\vec{\tau}}=\langle\Psi_{0}|c_{\mathbf{R}_{i}+\vec{\tau},\alpha}d_{\mathbf{R}_{i},\beta}-c_{\mathbf{R}_{i}+\vec{\tau},\beta}d_{\mathbf{R}_{i},\alpha}|\Psi_{0}\rangle.
\end{equation}
Here $\vec{\tau}_{m}$ (with $m=1,2,3$) denotes the three nearest neighbour bond directions of the A site. Therefore, the superconducting order parameter is given by
\begin{flalign}
\tilde{\Delta}_{SC,\alpha\beta}^{\vec{\tau}}&=\langle
c_{\mathbf{R}_{i}+\vec{\tau},\alpha}d_{\mathbf{R}_{i},\beta}-c_{\mathbf{R}_{i}+\vec{\tau},\beta}d_{\mathbf{R}_{i},\alpha}\rangle_{\Psi}
                     \nonumber\\
&=g_{t}^{SU(4)}(n,\nu_{0})\tilde{\Delta}_{\alpha\beta}^{\vec{\tau}}.
\end{flalign}
We use the parameter $\varphi$ to represent the phase difference between the pairing order parameters on the three nearest neighbour bonds, namely,
$\tilde{\Delta}_{\alpha\beta}^{\vec{\tau}_{2}}=\mathrm{e}^{i\varphi}\tilde{\Delta}_{\alpha\beta}^{\vec{\tau}_{1}}$ and
$\tilde{\Delta}_{\alpha\beta}^{\vec{\tau}_{3}}=\mathrm{e}^{i2\varphi}\tilde{\Delta}_{\alpha\beta}^{\vec{\tau}_{1}}$. $\varphi=2\pi/3$ yields the $d\pm id$ state and $\varphi=0$ yields
the extended $s$-wave state.

According to crystal field theory, in the two-dimensional hexagonal lattice there are two reasonable and favorable irreducible representations for even-parity
gap functions (recall that we only consider even pairing fields and here we set $k_{z}=0$). The first is the one-dimensional representation and its basis functions can be chosen as 1 or
$k_{x}^{2}+k_{y}^{2}$ which can yield the extended $s$-wave. The second is the two-dimensional representation and its basis functions consist of $k_{x}^{2}-k_{y}^{2}$ and $2k_{x}k_{y}$. The best equal weight combination of degenerate states $k_{x}^{2}-k_{y}^{2}$ and $2k_{x}k_{y}$ yields the $d\pm id$ state \cite{Sigrist1991,SchafferAndHonerkamp2014}. Our treatment is limited to zero temperature and thus we can obtain the self consistent equations for the order parameters
\begin{flalign}\label{eq:selfconsis_order_parameters}
&\tilde{\Delta}_{\alpha\beta}=\tilde{\Delta}_{\alpha\beta}^{\vec{\tau}_{1}},
                    \nonumber\\
&\chi=\sum_{\alpha}(\chi_{\langle i,j\rangle\alpha}+\chi_{\langle i,j\rangle\alpha}^{*})/2,
                          \nonumber\\
&n=\sum_{\alpha}\langle\Psi_{0}|d_{\mathbf{R}_{i},\alpha}^{\dagger}d_{\mathbf{R}_{i},\alpha}+c_{\mathbf{R}_{i}+\vec{\tau}_{1},\alpha}^{\dagger}c_{\mathbf{R}_{i}+\vec{\tau}_{1},\alpha}|\Psi_{0}\rangle/2,
\end{flalign}
by their expressions with respect to Bogoliubov transformation matrices $\hat{u}^{\mathbf{k}}$ and $\hat{v}^{\mathbf{k}}$. More details can be found in Appendix
\ref{appendixsec:BdGmethod}.
\begin{figure}
\centering
\includegraphics[width=0.48\textwidth]{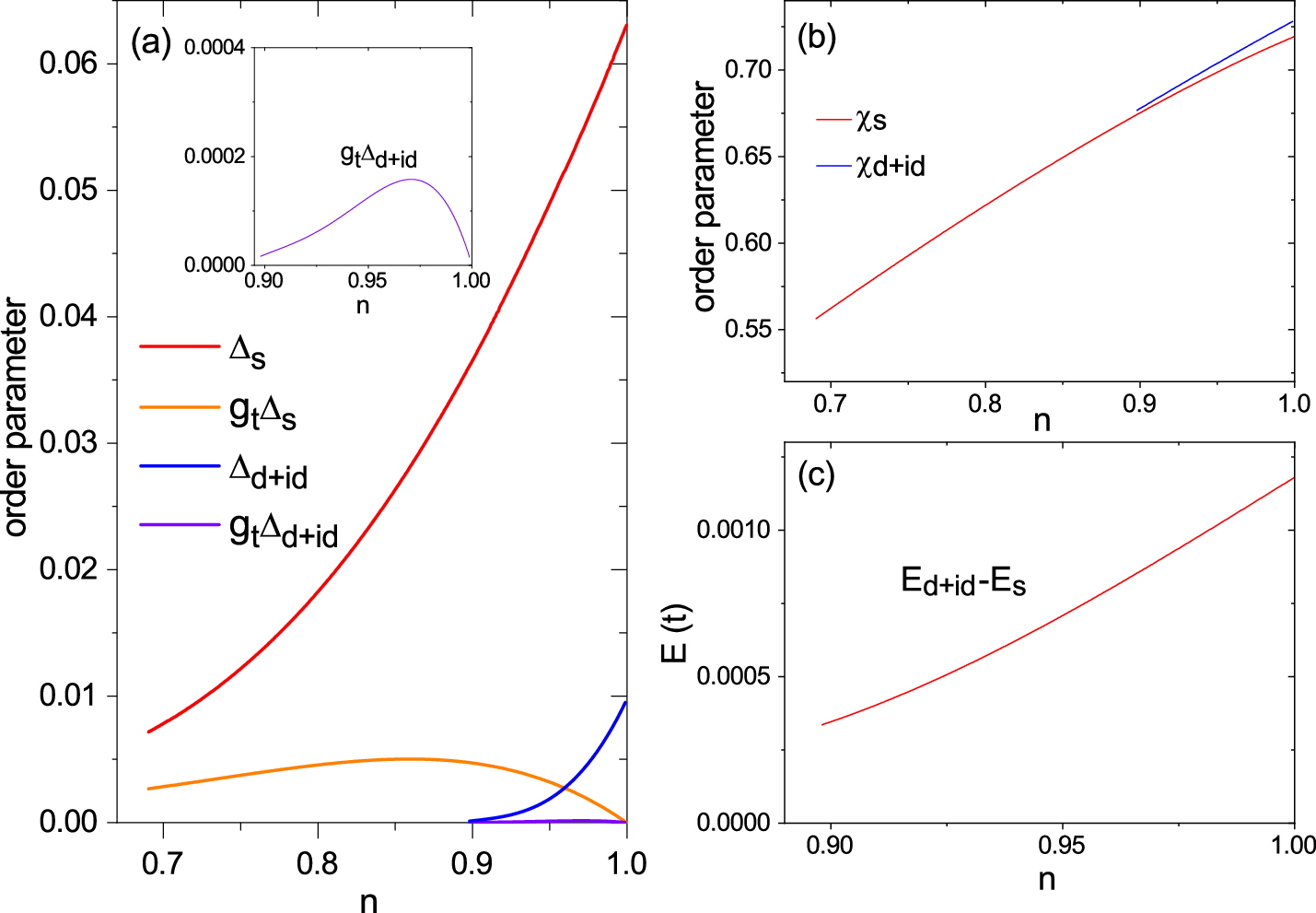}
\caption{\label{GapdiagramAtv1v2}The case for the extended $s$-wave state and the $d\pm id$ state scenarios below quarter filling. Here we choose $t=1$ and $J/t=0.3/8$. 
(a) The electron density dependence of flavor gaps, $\Delta_{s}$ and $\Delta_{d+id}$ (defined in Eq. (\ref{eq:selfconsis_order_parameters})), at zero temperature and superconducting
order parameters, $g_{t}\Delta_{s}$ and $g_{t}\Delta_{d+id}$ ($g_{t}$ is defined in Eq. (\ref{SU4gtfactor})), both their behaviors have the dome structure. (Inset) The enlargement of
that of the $d\pm id$ state for clarity. (b) The electron density dependence of kinetic energy order parameters, $\chi_{s}$ and $\chi_{d+id}$. (c) The difference in the total energy (per
site) between the $d\pm id$ state and the extended $s$-wave state, namely, $\Delta E=E_{d+id}-E_{s}$, where $E_{d+id}$ and $E_{s}$ are the total energies (given by Eq.
(\ref{RenormalazitionHvalue})) per site (in units of $t$) of the $d\pm id$ state and the extended $s$-wave state, respectively. }
\end{figure}

\begin{figure}
\centering
\includegraphics[width=0.48\textwidth]{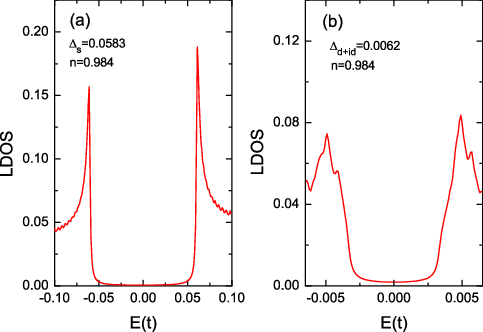}
\caption{\label{LDOSdiagramfine}LDOS versus E (in units of $t$). (a) The case of the extended $s$-wave state with the broadening factor $\eta=0.001t$ and
electron density $n=0.984$. (b) The case of the $d\pm id$ state with the broadening factor $\eta=0.0002t$ and electron density $n=0.984$. }
\end{figure}

\section{Numerical results}\label{sec:numericalresults}
Here we only concentrate on numerical results about the cases of hole doping away from integer filling number $\nu_{0}=1$ (corresponding to quarter filling) and $\nu_{0}=2$ (corresponding to half-filling). $t=1$ and $J/t=0.3/8$ are chosen for calculations. We obtained the extended $s$-wave state and the $d\pm id$ state, and here ``extended $s$-wave" and ``$d\pm id$" describe the orbital part of the pairing. In addition to $U(1)$ symmetry, the time reversal symmetry of the $d\pm id$ state is also broken and this is a prominent difference with respect to the
extended $s$-wave. In our numerical calculations we only found the flavor antisymmetric states $\tilde{\Delta}_{\alpha\beta}=-\tilde{\Delta}_{\beta\alpha}$ ($\alpha\ne\beta$) and this
can be easily understood from its definition. The states in this paper all belong to the flavor configuration:
$\tilde{\Delta}=\tilde{\Delta}_{12}=\tilde{\Delta}_{34}=\tilde{\Delta}_{14}=\tilde{\Delta}_{23}=\tilde{\Delta}_{13}=-\tilde{\Delta}_{24}$. This flavor configuration corresponds to
$\langle\overrightarrow{\Delta}_{\mathbf{R}_{i}+\vec{\tau}_{1},\mathbf{R}_{i}}\rangle_{0}=\left(0,2\tilde{\Delta},0,2i\tilde{\Delta},0,-2i\tilde{\Delta}\right)=\langle\tilde{c}_{i}^{t}\left(0,
i \sigma^{02}, 0, -\sigma^{23}, 0, \sigma^{21}\right) \tilde{c}_{j}\rangle_{0}$ which indicates the spin singlet pairing and the spin triplet pairing exist simultaneously in this flavor
configuration. More detail about the physical meaning of the flavor configuration can be found in Appendix \ref{appendixsec:flavorconfiguration}. As a matter of fact, let's make a
multiplication of phase factors to four kinds of electrons with different flavors, i.e., $c_{1}\to c_{1}^{\prime}=c_{1}\mathrm{e}^{i(\theta_{1}+\theta_{2}+2\theta_{4})/2}$, $c_{2}\to
c_{2}^{\prime}=c_{2}\mathrm{e}^{i(\theta_{1}-\theta_{3}+2\theta_{4})/2}$, $c_{3}\to c_{3}^{\prime}=c_{3}\mathrm{e}^{i(\theta_{2}-\theta_{3}+2\theta_{4})/2}$, and $c_{4}\to
c_{4}^{\prime}=c_{4}\mathrm{e}^{i\theta_{4}}$. Then we obtain $\tilde{\Delta}_{12}=\mathrm{e}^{-i\theta_{1}}\tilde{\Delta}_{34}$,
$\tilde{\Delta}_{14}=\mathrm{e}^{-i\theta_{3}}\tilde{\Delta}_{23}$, and $\tilde{\Delta}_{13}=-\mathrm{e}^{-i\theta_{2}}\tilde{\Delta}_{24}$. This indicates there are a large number of
degenerate states arising from the flavor phase degrees of freedom between different flavor pairing fields. These degenerate states can be transformed into each other by the gauge
transformation listed above, and therefore we only need to study one flavor configuration.

For the case of hole doping away from quarter filling, the electron density dependence of the pairing order parameter (RVB gap or flavor gap) is shown in Fig. \ref{GapdiagramAtv1v2}(a), and
here we denote $\Delta_{s}=\tilde{\Delta}_{12}$ and $\Delta_{d+id}=\tilde{\Delta}_{12}$ for the extended $s$-wave state and the $d\pm id$ state, respectively. The quantity $g_{t}\Delta$
describes the superconducting order parameter. As shown in Fig. \ref{GapdiagramAtv1v2}(a), in a whole doping region close to quarter filling, the magnitude of the pairing order parameter
and the superconducting order parameter of the extended $s$-wave state is substantially higher than that of the $d\pm id$ state. As a function of doping $\delta=1-n$, both of the
behaviors for $g_{t}\Delta$ of the extended $s$-wave state and of the $d\pm id$ state have a dome structure. $g_{t}\Delta$ vanishes linearly near $\delta=0$. The quantity $g_{t}\Delta$
can be understood as an approximate superconducting transition temperature $T_{c}$ \cite{Honerkamp2013,PWAnderson2004J.Phys.}. Their corresponding behaviors of the electron density
dependence of the kinetic energy order parameter are shown in Fig. \ref{GapdiagramAtv1v2}(b), and here we denote $\chi_{s}=\chi$ and $\chi_{d+id}=\chi$ for the extended $s$-wave state
and the $d\pm id$ state, respectively. The total energy of the extended $s$-wave state is evidently lower than that of the $d\pm id$ state in the whole doping region, as shown in Fig.
\ref{GapdiagramAtv1v2}(c). Here $E_{d+id}$ and $E_{s}$ are the total energies (given by Eq. (\ref{RenormalazitionHvalue})) per site of the $d\pm id$ state and the extended $s$-wave state,
respectively. Therefore, the extended $s$-wave state is the dominant pairing state in the case of hole doping away from quarter filling.

We also calculated the local density of states (LDOS) of the extended $s$-wave state and the $d\pm id$ state in the doping region close to the integer filling number $\nu_{0}=1$ as shown
in Fig. \ref{LDOSdiagramfine}. From the LDOS diagram it is obvious that both the extended $s$-wave state and the $d\pm id$ state are fully gapped. Here the LDOS is calculated by
$g_{t}^{SU(4)}\sum_{\alpha}A_{\alpha}^{i}(\omega)$, where $\langle
c^{\dagger}_{i,\alpha}c_{i,\alpha}\rangle_{\Psi_{0}}=\int_{-\infty}^{\infty}\frac{A_{\alpha}^{i}(\omega)}{\mathrm{e}^{\hbar\omega/\mathrm{k}_{B}T}+1}d\omega$.

\begin{figure}
\centering
\includegraphics[width=0.48\textwidth]{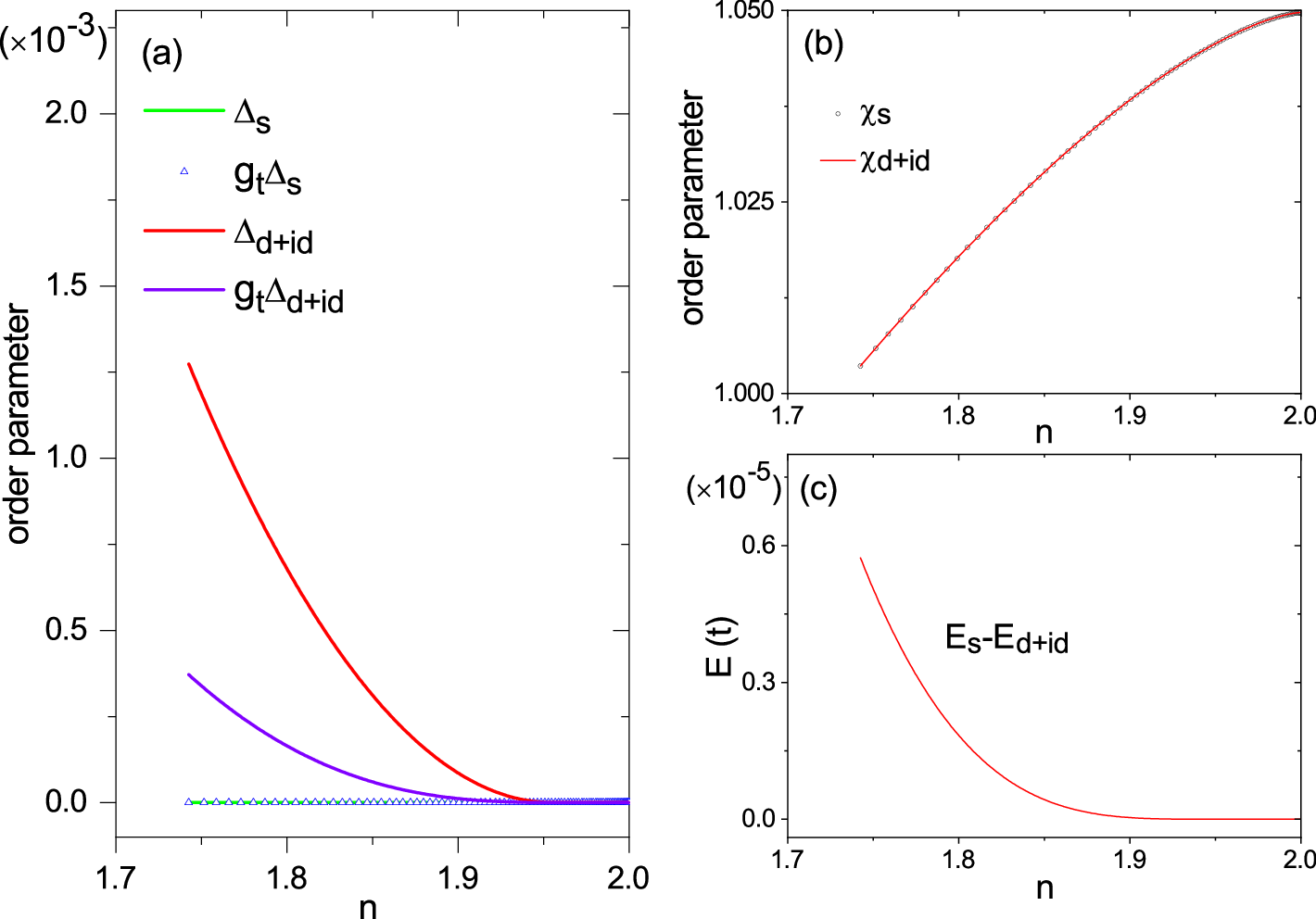}
\caption{\label{GapdiagramAtv2}The case for the extended $s$-wave state and the $d\pm id$ state scenarios below half-filling. (a) The electron density dependence of flavor gaps, $\Delta_{s}$ and $\Delta_{d+id}$ (defined in Eq. (\ref{eq:selfconsis_order_parameters})), at zero temperature and superconducting order parameters, $g_{t}\Delta_{s}$ and $g_{t}\Delta_{d+id}$ ($g_{t}$ is defined in Eq. (\ref{SU4gtfactor})). Note that the values of left axis need to be multiplied by $10^{-3}$. Here the values of $\Delta_{s}$ and of $g_{t}\Delta_{s}$ are exactly equal to zero in the given doping region, which corresponds to the u-RVB state \cite{Ogata_2008}. The values of $\Delta_{d+id}$ and of $g_{t}\Delta_{d+id}$ are exactly equal to zero in the region of $1.9552\le n\le 2$. (b) The kinetic energy order parameters behave identically for the extended $s$-wave state and the $d\pm id$ state. (c) The difference in the total energy (per site) between the extended $s$-wave state and the $d\pm id$ state, namely, $\Delta E=E_{s}-E_{d+id}$, where $E_{d+id}$ and $E_{s}$ are the total energies (given by Eq. (\ref{RenormalazitionHvalue})) per site (in units of $t$) of the $d\pm id$ state and the extended
$s$-wave state, respectively. Note that the values of left axis need to be multiplied by $10^{-5}$, and the values of $\Delta E$ are exactly equal to zero in the region of $1.9552\le n\le
2$ and positive below the doping region $n<1.9552$. Since the values of $\Delta E$ are negligible small, the total energy of the $d\pm id$ state is almost the same with that of the u-RVB
state.}
\end{figure}

For the case of hole doping away from half-filling, as shown in Fig. \ref{GapdiagramAtv2}(a), the values of the pairing order parameter and the superconducting order parameter of the extended
$s$-wave state are exactly equal to zero in the whole doping region presented in the figure, which corresponds to the uniform RVB (u-RVB) state \cite{Ogata_2008}. Therefore, the extended
$s$-wave state does not exist in this doping region. Since the values of $\Delta_{d+id}$ and of $g_{t}\Delta_{d+id}$ are exactly equal to zero in the region of $1.9552\le n\le 2$, the $d\pm
id$ state does not arise until doping more than $\delta=1-n\approx0.045$. However, note that the values of the left axis of Fig. \ref{GapdiagramAtv2}(a) need to be multiplied by a factor $10^{-3}$ and the maximum magnitude of $\Delta_{d+id}$ is around $10^{-3}$ in the presented doping region. Thus the values of $\Delta_{d+id}$ are negligible in this doping region. Note that the magnitudes of $\Delta_{d+id}$ in this doping region are also much smaller than that of $\Delta_{d+id}$ in the case of doping region close to quarter filling. Therefore, the superconductivity in this doping region can be neglected. The corresponding kinetic energy order parameters, $\chi_{s}$ and $\chi_{d+id}$, behave identically as shown in Fig. \ref{GapdiagramAtv2}(b). The total energy of the $d\pm id$ state is slightly lower than that of the u-RVB state, as shown in Fig. \ref{GapdiagramAtv2}(c). Therefore, near half-filling, the $d\pm id$ state is more favorable in the doping region where $\Delta_{d+id}$ is non-vanishing. Moreover, It can be inferred, from the negligible values of $\Delta_{d+id}$ and $\Delta_{s}=0$ near half-filling, that the pairing fields are unstable in this doping region.

\section{Discussion and conclusion}\label{sec:discusionandconclusion}
Since there is a considerable overlap between the nodes of the $d\pm id$ form factor and the Fermi surface near quarter filling \cite{Honerkamp2013}, one could anticipate the
free energy of the $d\pm id$ state will be higher than that of the extended $s$-wave state. This may explain that in a whole doping region close to quarter filling, the extended $s$-wave state is
much more favorable than the $d\pm id$ state. Moreover, there is almost no superconductivity in the doping range close to half-filling, in contrast with the SU(2) case where the $d\pm id$ state is more favorable than the extended $s$-wave state close to half-filling \cite{Honerkamp2013}. The flavor fluctuation effect in the SU(4) $t$-$J$ model being more substantial than the SU(2) case and the low electron density close to half-filling should be the reasons. On the other hand, the pairing field $\overrightarrow{\Delta}_{i j}$ is formed by four fermions rather than two fermions, which also makes the pairing fields unstable in the low filling region.

Next we discuss the relevance of our results to TBG. For TBG, the microscopic two-orbital extended Hubbard model on the emergent honeycomb lattice \cite{LiangFuPRB2018}, constructed from
band structure calculations projected to low-energy bands and the analysis of energy bands at all high symmetry points, is more complicated than the simplest two-orbital honeycomb
Hubbard model studied here. Our model does not include some terms of band structure details and pair-hopping interaction terms. The simplified model used here might provide a
reference point for understanding the correlated electron physics in TBG. Due to TBG's similarity to the cuprates, the large $U$ regime of the SU(4) honeycomb Hubbard model is important for TBG.

Based on the real-space density matrix renormalization group (DMRG) simulation, the metal-insulator transition around $U_{c}/t=$2.5--3 and a nonmagnetic Mott insulator in the large
$U$ regime are identified in the two-orbital Hubbard model on the honeycomb lattice (i.e., Eq. (\ref{eq: su(4) Hubbard model})) at quarter filling. Moreover, the featureless nonmagnetic Mott
insulator has no signs of charge and spin-orbital density wave fluctuations or orders \cite{D.N.Sheng2019}. Corresponding to $J/t=0.3/8$ used in our work, $U/t=20/3$ is consistent with
the large $U$ regime. At quarter filling, the RVB state as the trial wavefunction in our work is also featureless nonmagnetic state.

The RVB state is of rare occurrence in Mott insulators. However, the RVB superconducting state may be realized by introducing a finite number of holes into the Mott insulator \cite{B.Edegger}. The pairing of the RVB state directly originates from the superexchange of the Mott insulator. As shown in Fig. \ref{GapdiagramAtv1v2}(a),
the feature of the superconducting dome in our results in the vicinity of quarter filling is consistent with that of the superconductivity observed in TBG \cite{Y.CaoSu}. Corresponding
to the region around the charge neutrality point of TBG, the superconducting state does not exist in the case of hole doping away from half-filling, which is also consistent with TBG experiments. Therefore, the scenario of the RVB state might provide an appropriate description for the states observed in TBG.

For comparing with other studies of the pairing in TBG, we remark that the spin-singlet and spin-triplet Cooper pairs are not sharply defined for our extended $s$-wave and $d\pm id$
pairing due to the SU(4) symmetry. In TBG, it is important to consider the effects of SU(4) symmetry-breaking perturbations on the honeycomb Hubbard model. We assume that the dominant
interaction is the Hund's coupling $-V\sum_{i}(\mathbf{S}_{i})^{2}$, where $V>0$ and $\mathbf{S}_{i}$ denotes the total spin on site $i$. Then the previous three components of
$\overrightarrow{\Delta}_{i j}$ defined in Eq. (\ref{eq:pairingform}) (i.e., the spin-triplet and orbital-singlet pairing field) is more favorable \cite{CenkeXu}. Consequently, the
extended $s$-wave pairing and the $d\pm id$ pairing obtained in our study are more favorable in spin-triplet and orbital-singlet. The parity of the pair can also exchange the
two orbitals (valleys) of TBG, and thus, in the large microscopic Brillouin zone, the even-parity extended $s$-wave state and the even-parity $d\pm id$ state become the odd-parity
$p$-wave state and the odd-parity $f\pm if$ state, respectively, according to orbital-singlet. Therefore, we infer that the spin-triplet $p$-wave superconductivity is dominant in the
vicinity of quarter-filling (two electrons per superlattice unit cell) in TBG, according to Fig. \ref{GapdiagramAtv1v2}. This is different from the results of Ref. \cite{CenkeXu}, where
the spin-triplet $f\pm if$ pairing in the large Brillouin zone, is proposed for TBG by the SU(4) $t$-$J$ model on the triangular lattice near half-filling, with the
Hund's coupling. Our spin-triplet $p$-wave pairing is similar to that in Ref. \cite{PhysRevLett.122.026801}, where the superconductivity in TBG is explained as a consequence of the
Kohn-Luttinger instability. In Ref. \cite{PhysRevB.98.241407}, based on the functional renormalization group (FRG) techniques, the $d\pm id$ pairing near quarter-filling is found on the
simplest two-orbital honeycomb Hubbard model (i.e, Eq. (\ref{eq: su(4) Hubbard model})) at the moderate $U/t=2$. Their $d\pm id$ pairing occurs between ($\uparrow$, $p_{x}$ orbital,
lower band) and ($\downarrow$, $p_{y}$ orbital, lower band), and this restriction of the pairing between particles with opposing quantum numbers does not exist in our study. This state corresponds to a single component of the $d\pm id$ pairing in the mini Brillouin zone with flavor structure near quarter-filling in our study. In Ref. \cite{PhysRevB.99.094521}, a
spin-triplet and orbital-singlet $f$-wave pairing is found on the two-orbital extended Hubbard model on the honeycomb lattice near quarter-filling from weak to moderate coupling, which
introduces a warping hopping term with the $f$-wave factor and the Hund's coupling term to break the SU(4) symmetry.

The Gutzwiller approximation adopted in our present study is a conventional method to treat the strong correlation problem. The homogeneous electron density and the flavor symmetry are assumed. For studying the charge density wave stability, the spin-orbital density wave stability, and the local effects due to vortex cores and impurities, the more complicated renormalization
factors derived from the case of the inhomogeneous particle density, are much demanded. Furthermore, the more complicated two-orbital extended Hubbard model for TBG can be
studied by this technique. We leave these studies to future work.

In summary, we obtain the Gutzwiller renormalization factors of the SU(4) $t$-$J$ model and the corresponding renormalized Hamiltonian. Utilizing the RVB state as a trial wave function and
the mean-field approximation to this renormalized Hamiltonian on the honeycomb lattice case, we find that the extended $s$-wave state is much more favorable than the $d\pm id$ state in
the doping region close to quarter filling, and the superconductivity almost disappears in the doping region close to half-filling. The spin-singlet pairing and the spin-triplet pairing can
exist simultaneously in the pairing state of the SU(4) case.

\begin{acknowledgments}
We thank Fu-Chun Zhang, Jian Kang, and Ting-Kuo Lee for helpful discussions. This work is supported by the National Key Research and Development Program of China (Grants Nos.
2017YFA0304204 and 2016YFA0300504), the National Natural Science Foundation of China Grant No. 11625416, and the Shanghai Municipal Government (Grants Nos. 19XD1400700 and 19JC1412702).

\end{acknowledgments}

\appendix

\section{The Bogoliubov-de Gennes method}\label{appendixsec:BdGmethod}
The expectation value of the renormalized Hamiltonian is
\begin{flalign}\label{RenormalazitionHvalue}
\langle \Psi_{0}|H^{\prime}|\Psi_{0}\rangle=&-tg_{t}^{SU(4)}\sum_{\langle i,j\rangle}\sum_{\alpha=1}^{4}(\chi_{\langle i,j\rangle\alpha}+\mathrm{h.c.})
            \nonumber\\
&-\frac{5}{4}Jg_{S}^{SU(4)}\sum_{\langle i,j\rangle}\left(\langle\overrightarrow{\Delta}_{i j}\rangle_{0}\right)^{\dagger} \cdot \langle\overrightarrow{\Delta}_{i j}\rangle_{0}
            \nonumber\\
&-\frac{15}{4}Jg_{S}^{SU(4)}\sum_{\langle i,j\rangle}\left(\sum_{\alpha}\chi_{\langle i,j\rangle \alpha}^{*}\sum_{\beta}\chi_{\langle i,j\rangle\beta}\right).
\end{flalign}
We minimize the energy under the constraints $\sum_{i}\langle \Psi_{0}|\hat{n}_{i}|\Psi_{0}\rangle=N_{e}$, $\langle \Psi_{0}|\Psi_{0}\rangle=1$, where $N_{e}$ is the total number of
electrons. Namely, minimizing the function $W=\langle \Psi_{0}|H^{\prime}|\Psi_{0}\rangle -\lambda(\langle \Psi_{0}|\Psi_{0}\rangle-1)-\mu(\sum_{i}\langle
\Psi_{0}|\hat{n}_{i}|\Psi_{0}\rangle-N_{e})$ which yields the variational relation
\begin{flalign}\label{variationalrelation}
0=&\frac{\delta W}{\delta\left\langle\Psi_{0} |\right.}
      \nonumber\\
=&\sum_{\langle i, j\rangle, \alpha} \frac{\partial W}{\partial \chi_{\langle i, j\rangle, \alpha}} \frac{\delta \chi_{\langle i, j\rangle, \alpha}}{\delta\left\langle\Psi_{0}
|\right.}+\mathrm{h.c.}
      \nonumber\\
&+\sum_{\langle i, j\rangle, \alpha\beta} \frac{\partial W}{\partial \Delta_{\langle i, j\rangle, \alpha\beta}} \frac{\delta \Delta_{\langle i, j\rangle,
\alpha\beta}}{\delta\left\langle\Psi_{0} |\right.}+\mathrm{h.c.}
      \nonumber\\
&+\sum_{i} \frac{\partial W}{\partial n_{i}} \frac{\delta n_{i}}{\delta\left\langle\Psi_{0}| \right.}-\lambda | \Psi_{0} \rangle,
\end{flalign}
where $n_{i}=\langle\Psi_{0}|\hat{n}_{i}|\Psi_{0}\rangle$. For an operator $\hat{O}$, $\delta\langle\Psi_{0}|\hat{O}| \Psi_{0}\rangle / \delta\langle\Psi_{0}|=\hat{O}| \Psi_{0}\rangle$.
Therefore we obtain the Schr$\ddot{\mathrm{o}}$dinger equation $H_{\mathrm{MF}}| \Psi_{0} \rangle =\lambda | \Psi_{0} \rangle$ and
\begin{flalign}\label{RMHamiltonian1}
H_{\mathrm{MF}}=&\sum_{\langle i, j\rangle, \alpha} \frac{\partial W}{\partial \chi_{\langle i, j\rangle, \alpha}} c_{i, \alpha}^{\dagger} c_{j, \alpha}+\mathrm{h.c.}
                  \nonumber\\
&+\sum_{\langle i, j\rangle, \alpha\beta}^{\alpha\ne\beta} \frac{\partial W}{\partial \Delta_{\langle i, j\rangle, \alpha\beta}}  c_{i, \alpha} c_{j, \beta}+\mathrm{h.c.}+\sum_{i,
\alpha} \frac{\partial W}{\partial n_{i}} \hat{n}_{i, \alpha}.
\end{flalign}
Here
\begin{equation}\label{partialn}
\frac{\partial W}{\partial n_{i}}=\frac{\partial W}{\partial n}=-\mu+\left[\frac{\partial \langle\Psi_{0}|H^{\prime}|\Psi_{0}\rangle}{\partial n}\right]_{g},
\end{equation}
\begin{equation}\label{partialwchi}
\frac{\partial W}{\partial \chi_{\langle i, j\rangle,a}}=-tg_{t}^{SU(4)}-\frac{15}{4}Jg_{S}^{SU(4)}\sum_{\alpha}\chi^{*}_{\langle i,j\rangle ,\alpha},
\end{equation}
\begin{equation}\label{partialwdelta}
\frac{\partial W}{\partial \Delta_{\langle i, j\rangle,\alpha\beta}}=\frac{5}{2}Jg_{S}^{SU(4)}(\Delta_{\langle i,j\rangle\beta\alpha}^{*}-\Delta_{\langle i,j\rangle\alpha\beta}^{*}),
\end{equation}
where $\left[\frac{\partial \langle\Psi_{0}|H^{\prime}|\Psi_{0}\rangle}{\partial n}\right]_{g}$ in Eq. (\ref{partialn}) being the derivate of $\langle\Psi_{0}|H^{\prime}|\Psi_{0}\rangle$
with respect to $n$ via the renormalization g-factors and $\alpha\ne \beta$ in Eq. (\ref{partialwdelta}). $-\partial W/\partial n$ can be viewed as the effective chemical potential and
we let $\tilde{\mu}=-\partial W/\partial n$. For simplicity, in this paper we only consider real $\chi_{\langle i,j\rangle ,\alpha}$. On the honeycomb lattice, for convenience, Eq.
(\ref{RMHamiltonian1}) can be rewritten as
\begin{flalign}\label{RMHamiltonian2}
H_{\mathrm{MF}}=&\sum_{\langle i, j\rangle, \alpha} \frac{\partial W}{\partial \chi_{\langle i, j\rangle, \alpha}} c_{i, \alpha}^{\dagger} d_{j, \alpha}+\mathrm{h.c.}
                  \nonumber\\
&+\sum_{\langle i, j\rangle, \alpha\beta}^{\alpha\ne\beta} \frac{\partial W}{\partial \Delta_{\langle i, j\rangle, \alpha\beta}}  c_{i, \alpha} d_{j,
\beta}+\mathrm{h.c.}-\tilde{\mu}\sum_{i, \alpha}\hat{n}_{i, \alpha}.
\end{flalign}
Here we denote by $d$ and $c$ the electron annihilation operators related with the A and B sublattices of the honeycomb lattice, respectively. We can Fourier-transform the Hamiltonian and
diagonalize the kinetic part by using the trick \cite{Honerkamp2013} as follows:
\begin{flalign}\label{eq:Honercamptrick}
&c_{\mathbf{k}\alpha}=\frac{1}{\sqrt{2}}(f_{\mathbf{k}\alpha}+g_{\mathbf{k}\alpha}),
    \nonumber\\
&d_{\mathbf{k}\alpha}=\frac{1}{\sqrt{2}}\mathrm{e}^{-i\phi_{\mathbf{k}}}(f_{\mathbf{k}\alpha}-g_{\mathbf{k}\alpha}),
\end{flalign}
where $\phi_{\mathbf{k}}=\arg(\gamma_{\mathbf{k}})$ and $\gamma_{\mathbf{k}}=\sum_{m=1}^{3}\mathrm{e}^{-i\mathbf{k}\cdot \vec{\tau}_{m}}$. Here $\vec{\tau}_{m}$, with $m=1,2,3$, denotes
the three nearest neighbour bond directions of the A site. Finally we obtain
\begin{flalign}\label{eq:RMHamiltonian3}
H_{\mathrm{MF}}=\sum_{\mathbf{k}}H_{\mathbf{k}}-\sum_{\mathbf{k}}\sum_{\alpha}2\tilde{\mu},
\end{flalign}
with
\begin{flalign}\label{eq:kHamiltonian}
&H_{\mathbf{k}}=\sum_{\alpha}\frac{1}{2}[(\epsilon_{\mathbf{k}}-\tilde{\mu})f_{\mathbf{k}\alpha}^{\dagger}f_{\mathbf{k}\alpha}+(-\epsilon_{\mathbf{k}}-\tilde{\mu})g_{\mathbf{k}\alpha}^{\dagger}g_{\mathbf{k}\alpha}]
                 \nonumber\\
&-\sum_{\alpha}\frac{1}{2}[(\epsilon_{\mathbf{k}}-\tilde{\mu})f_{-\mathbf{k}\alpha}f_{-\mathbf{k}\alpha}^{\dagger}+(-\epsilon_{\mathbf{k}}-\tilde{\mu})g_{-\mathbf{k}\alpha}g_{-\mathbf{k}\alpha}^{\dagger}]
     \nonumber\\
&-\frac{5}{4}Jg_{S}^{SU(4)}\sum_{\alpha\beta}^{\alpha\ne\beta}[\Delta_{\mathbf{k},\alpha\beta}^{i}(f_{\mathbf{k}\alpha}^{\dagger}f_{-\mathbf{k}\beta}^{\dagger}-g_{\mathbf{k},\alpha}^{\dagger}g_{-\mathbf{k}\beta}^{\dagger})+\mathrm{h.c.}]
                  \nonumber\\
&-\frac{5}{4}Jg_{S}^{SU(4)}\sum_{\alpha\beta}^{\alpha\ne\beta}[\Delta_{\mathbf{k},\alpha\beta}^{I}(f_{\mathbf{k}\alpha}^{\dagger}g_{-\mathbf{k}\beta}^{\dagger}-g_{\mathbf{k},\alpha}^{\dagger}f_{-\mathbf{k}\beta}^{\dagger})+\mathrm{h.c.}],
\end{flalign}
where $\epsilon_{\mathbf{k}}=|\gamma_{\mathbf{k}}|\partial W/\partial \chi_{\langle i,j\rangle,a}$ and
\begin{flalign}\label{intrainterbandpairing}
&\Delta_{\mathbf{k},\alpha\beta}^{i}=-\tilde{\Delta}_{\alpha\beta}\Gamma_{\mathbf{k}}^{i},
     \nonumber\\
&\Delta_{\mathbf{k},\alpha\beta}^{I}=-\tilde{\Delta}_{\alpha\beta}\Gamma_{\mathbf{k}}^{I},
     \nonumber\\
&\Gamma_{\mathbf{k}}^{i}=\sum_{m=1}^{3}\mathrm{e}^{i\varphi(m-1)}\cos(\mathbf{k}\cdot\vec{\tau}_{m}+\phi_{\mathbf{k}}),
     \nonumber\\
&\Gamma_{\mathbf{k}}^{I}=\sum_{m=1}^{3}\mathrm{e}^{i\varphi(m-1)}i\sin(\mathbf{k}\cdot\vec{\tau}_{m}+\phi_{\mathbf{k}}),
     \nonumber\\
&\tilde{\Delta}_{\alpha\beta}=\tilde{\Delta}_{\alpha\beta}^{\vec{\tau}_{1}},
     \nonumber\\
&\tilde{\Delta}_{\alpha\beta}^{\vec{\tau}}=\langle\Psi_{0}|c_{\mathbf{R}_{i}+\vec{\tau},\alpha}d_{\mathbf{R}_{i},\beta}-c_{\mathbf{R}_{i}+\vec{\tau},\beta}d_{\mathbf{R}_{i},\alpha}|\Psi_{0}\rangle.
\end{flalign}
Here $\varphi$ represents the phase difference between the pairing order parameters in the three nearest neighbour bond directions, namely,
$\tilde{\Delta}_{\alpha\beta}^{\vec{\tau}_{2}}=\mathrm{e}^{i\varphi}\tilde{\Delta}_{\alpha\beta}^{\vec{\tau}_{1}}$ and
$\tilde{\Delta}_{\alpha\beta}^{\vec{\tau}_{3}}=\mathrm{e}^{i2\varphi}\tilde{\Delta}_{\alpha\beta}^{\vec{\tau}_{1}}$. According to crystal field theory, in the two-dimensional hexagonal
lattice there are only the $d\pm id$ state and extended $s$-wave state which are reasonable and favorable \cite{Sigrist1991,SchafferAndHonerkamp2014}. $\varphi=2\pi/3$ yields the $d\pm
id$ state and $\varphi=0$ yields the extended $s$-wave state. By using the sixteen-component notation
$\tilde{f}_{\mathbf{k}}=(\tilde{f}_{\mathbf{k}\alpha},\tilde{g}_{\mathbf{k}\alpha},\tilde{f}^{\dagger}_{-\mathbf{k}\alpha},\tilde{g}^{\dagger}_{-\mathbf{k}\alpha})^{T}$ (for convenience
here $\tilde{f}_{\mathbf{k}\alpha}=(f_{\mathbf{k}1},f_{\mathbf{k}2},f_{\mathbf{k}3},f_{\mathbf{k}4})$ and
$\tilde{g}^{\dagger}_{-\mathbf{k}\alpha}=(g^{\dagger}_{-\mathbf{k}1},g^{\dagger}_{-\mathbf{k}2},g^{\dagger}_{-\mathbf{k}3},g^{\dagger}_{-\mathbf{k}4})$) we can rewrite $H_{\mathbf{k}}$
as
\begin{flalign}\label{matrixHk}
H_{\mathbf{k}}=\tilde{f}_{\mathbf{k}}^{\dagger}\hat{\mathscr{E}}_{\mathbf{k}}\tilde{f}_{\mathbf{k}},
\end{flalign}
and
\begin{flalign}\label{matrixcore}
&\hat{\mathscr{E}}_{\mathbf{k}}=\left[ \begin{array}{cc}
\hat{\varepsilon}_{\mathbf{k}}/2&\hat{\Delta}_{\mathbf{k}}\\
\hat{\Delta}_{\mathbf{k}}^{\dagger}&-\hat{\varepsilon}_{\mathbf{k}}/2
\end{array} \right],
\end{flalign}
where
\begin{flalign}\label{subcorematrix}
&\hat{\varepsilon}_{\mathbf{k}}=\left[ \begin{array}{cc}
(\epsilon_{\mathbf{k}}-\tilde{\mu})\mathbf{1}_{4\times4}&0\\
0&(-\epsilon_{\mathbf{k}}-\tilde{\mu})\mathbf{1}_{4\times4}\end{array} \right],
    \nonumber\\
&\hat{\Delta}_{\mathbf{k}}=\left[ \begin{array}{cc}
\hat{L}_{\mathbf{k}}&\hat{R}_{\mathbf{k}}\\
-\hat{R}_{\mathbf{k}}&-\hat{L}_{\mathbf{k}}\end{array} \right],
    \nonumber\\
&(\hat{L}_{\mathbf{k}})_{\alpha\beta}=\left\{ \begin{array}{ll}-\frac{5}{4}Jg_{S}^{SU(4)}\Delta_{\mathbf{k},\alpha\beta}^{i}&(\alpha\ne\beta)\\ 0&(\alpha=\beta)\end{array}\right. ,
    \nonumber\\
&(\hat{R}_{\mathbf{k}})_{\alpha\beta}=\left\{ \begin{array}{ll}-\frac{5}{4}Jg_{S}^{SU(4)}\Delta_{\mathbf{k},\alpha\beta}^{I}&(\alpha\ne\beta)\\ 0&(\alpha=\beta)\end{array}\right. .
\end{flalign}
Note that we have the relation $\hat{\Delta}_{\mathbf{k}}=-\hat{\Delta}_{-\mathbf{k}}^{T}$. We can diagonalize $H_{\mathbf{k}}$ as follows:
\begin{equation}\label{eq:kHamiltonian2}
H_{\mathbf{k}}=\sum_{n=1}^{8}(E_{\mathbf{k}n}a_{\mathbf{k}n}^{\dagger}a_{\mathbf{k}n}-E_{-\mathbf{k}n}a_{-\mathbf{k}n}a_{-\mathbf{k}n}^{\dagger}),
\end{equation}
by the (unitary) Bogoliubov transformation \cite{Sigrist1991},
\begin{flalign}\label{Bogoliubovtransformation}
&f_{\mathbf{k}\alpha}=\sum_{n}^{\prime}\left(u_{f\alpha,n}^{\mathbf{k}}a_{\mathbf{k}n}+(v_{f\alpha,n}^{-\mathbf{k}})^{*}a_{-\mathbf{k}n}^{\dagger}\right),
          \nonumber\\
&g_{\mathbf{k}\alpha}=\sum_{n}^{\prime}\left(u_{g\alpha,n}^{\mathbf{k}}a_{\mathbf{k}n}+(v_{g\alpha,n}^{-\mathbf{k}})^{*}a_{-\mathbf{k}n}^{\dagger}\right),
\end{flalign}
where $a_{\mathbf{k}n}$ and $a_{\mathbf{k}n}^{\dagger}$ (n=1,2,3,$\cdots$,8) satisfy the anticommutation relations of fermions ($\{
a_{\mathbf{k}n},a_{\mathbf{k}^{\prime}m}^{\dagger}\}=\delta_{\mathbf{k}\mathbf{k}^{\prime}}\delta_{nm}$, $\{a_{\mathbf{k}n},a_{\mathbf{k}^{\prime}m}\}=0$) and generate the elementary
excitations of the system. The prime sign above summation implies only those states with positive energy are counted. With the help of the sixteen-component notation
$\mathbf{a}_{\mathbf{k}}=(a_{\mathbf{k}1},\cdots,a_{\mathbf{k}8},a_{-\mathbf{k}1}^{\dagger},\cdots,a_{-\mathbf{k}8}^{\dagger})^{T}$ we can obtain a more compact formulation of Eq.
(\ref{Bogoliubovtransformation}): $\tilde{f}_{\mathbf{k}}=U_{\mathbf{k}}\mathbf{a}_{\mathbf{k}}$ with
\begin{flalign}\label{unitarybogo}
U_{\mathbf{k}}=\left[ \begin{array}{cc}
\hat{u}^{\mathbf{k}}&(\hat{v}^{-\mathbf{k}})^{*}\\
\hat{v}^{\mathbf{k}}&(\hat{u}^{-\mathbf{k}})^{*}\end{array} \right]\quad \textrm{and}\quad U_{\mathbf{k}}U_{\mathbf{k}}^{\dagger}=\mathbf{1}.
\end{flalign}
Here $8\times8$ Bogoliubov transformation matrices $\hat{u}^{\mathbf{k}}$ and $\hat{v}^{\mathbf{k}}$ are defined by Eq. (\ref{Bogoliubovtransformation}). Using this formalism, we rewrite
the Eq. (\ref{eq:kHamiltonian2}) as
\begin{equation}\label{eq:kHamiltonian3}
U_{\mathbf{k}}^{\dagger}\hat{\mathscr{E}}_{\mathbf{k}}U_{\mathbf{k}}=\hat{E}_{\mathbf{k}},
\end{equation}
where $\hat{E}_{\mathbf{k}}=\textrm{diag}(E_{\mathbf{k}1},\cdots,E_{\mathbf{k}8},-E_{-\mathbf{k}1},\cdots,-E_{-\mathbf{k}8})$. Therefore we obtain the Bogoliubov-de Gennes (BdG)
\cite{P.G.deGennes1966} equations as follows:
\begin{flalign}\label{eq:kBdGEq}
\hat{\mathscr{E}}_{\mathbf{k}}\left( \begin{array}{c}\underline{u_{\cdot n}^{\mathbf{k}}} \\ \\ \underline{v_{\cdot n}^{\mathbf{k}}}\end{array} \right)=E_{\mathbf{k}n}\left(
\begin{array}{c}\underline{u_{\cdot n}^{\mathbf{k}}} \\ \\ \underline{v_{\cdot n}^{\mathbf{k}}}\end{array}\right).
\end{flalign}
Here $\underline{u_{\cdot n}^{\mathbf{k}}}$($\underline{v_{\cdot n}^{\mathbf{k}}}$) is the $n$-th column of $\hat{u}^{\mathbf{k}}$($\hat{v}^{\mathbf{k}}$). From Eq. (\ref{eq:kBdGEq}) we
can know that if $\underline{u_{\cdot n}^{\mathbf{k}}},\underline{v_{\cdot n}^{\mathbf{k}}},E_{\mathbf{k}n}$ is the solution then $(\underline{v_{\cdot
n}^{-\mathbf{k}}})^{*},(\underline{u_{\cdot n}^{-\mathbf{k}}})^{*},-E_{-\mathbf{k}n}$ is also the solution. This property is consistent with the unitary Bogoliubov transformation Eq.
(\ref{Bogoliubovtransformation}). Note that
$\langle\Psi_{0}|a_{\mathbf{k}n}^{\dagger}a_{\mathbf{k}^{\prime}m}|\Psi_{0}\rangle=\delta_{\mathbf{k}\mathbf{k}^{\prime}}\delta_{nm}f(2E_{\mathbf{k}n})$,
$\langle\Psi_{0}|a_{\mathbf{k}n}a_{\mathbf{k}m}|\Psi_{0}\rangle=0$ and $f(2E_{\mathbf{k}n})=(\mathrm{exp}(2E_{\mathbf{k}n}/k_{B}T)+1)^{-1}$. Our treatment is limited to zero temperature
and thus we obtain the self consistent equations for the order parameters
\begin{widetext}
\begin{flalign}\label{eq:selfconsistenteqs}
\tilde{\Delta}_{\alpha\beta}=&(2N_{s})^{-1}\sum_{\mathbf{k}}\mathrm{e}^{i(\phi_{\mathbf{k}}+\mathbf{k}\cdot\vec{\tau}_{1})}\sum_{n}^{\prime}[u_{f\alpha,n}^{\mathbf{k}}(v_{f\beta,n}^{\mathbf{k}})^{*}
-u_{f\beta,n}^{\mathbf{k}}(v_{f\alpha,n}^{\mathbf{k}})^{*}-u_{g\alpha,n}^{\mathbf{k}}(v_{g\beta,n}^{\mathbf{k}})^{*}+u_{g\beta,n}^{\mathbf{k}}(v_{g\alpha,n}^{\mathbf{k}})^{*}
     \nonumber\\
&-u_{f\alpha,n}^{\mathbf{k}}(v_{g\beta,n}^{\mathbf{k}})^{*}+u_{g\alpha,n}^{\mathbf{k}}(v_{f\beta,n}^{\mathbf{k}})^{*}+u_{f\beta,n}^{\mathbf{k}}(v_{g\alpha,n}^{\mathbf{k}})^{*}-u_{g\beta,n}^{\mathbf{k}}(v_{f\alpha,n}^{\mathbf{k}})^{*}],\quad\quad(\alpha\ne\beta)
     \nonumber\\
\chi=&(2N_{s})^{-1}\sum_{\alpha}\sum_{\mathbf{k}}\frac{1}{2}\mathrm{e}^{-i(\phi_{\mathbf{k}}+\mathbf{k}\cdot\vec{\tau}_{1})}\sum_{n}^{\prime}[|v_{f\alpha,n}^{-\mathbf{k}}|^{2}-|v_{g\alpha,n}^{-\mathbf{k}}|^{2}+(v_{f\alpha,n}^{-\mathbf{k}})^{*}v_{g\alpha,n}^{-\mathbf{k}}-v_{f\alpha,n}^{-\mathbf{k}}(v_{g\alpha,n}^{-\mathbf{k}})^{*}+\mathrm{h.c.}],
     \nonumber\\
n=&(2N_{s})^{-1}\sum_{\alpha}\sum_{\mathbf{k}}\sum_{n}^{\prime}(|v_{f\alpha,n}^{-\mathbf{k}}|^{2}+|v_{g\alpha,n}^{-\mathbf{k}}|^{2}),
\end{flalign}
\end{widetext}
where $2N_{s}$ is the total number of sites. $\chi=\sum_{\alpha}(\chi_{\langle i,j\rangle\alpha}+\chi_{\langle i,j\rangle\alpha}^{*})/2$ and
$n=\sum_{\alpha}\langle\Psi_{0}|d_{\mathbf{R}_{i},\alpha}^{\dagger}d_{\mathbf{R}_{i},\alpha}+c_{\mathbf{R}_{i}+\vec{\tau}_{1},\alpha}^{\dagger}c_{\mathbf{R}_{i}+\vec{\tau}_{1},\alpha}|\Psi_{0}\rangle/2$.

\section{Pairing field flavor configurations of states}\label{appendixsec:flavorconfiguration}
About the pairing field we have the relation
\begin{flalign}\label{eq:pairing field relation}
\overrightarrow{\Delta}_{i j}=&\tilde{c}_{i}^{t}\left(-\sigma^{32}, i \sigma^{02}, \sigma^{12}, -\sigma^{23}, i\sigma^{20}, \sigma^{21}\right) \tilde{c}_{j}
                                 \nonumber\\
=&\tilde{c}_{i}^{t}\left(i\hat{\sigma}\sigma^{2}\otimes\sigma^{2},\sigma^{2}\otimes i\hat{\sigma}\sigma^{2}\right) \tilde{c}_{j},
\end{flalign}
where $\hat{\sigma}=\sigma^{1}\hat{e}_{1}+\sigma^{2}\hat{e}_{2}+\sigma^{3}\hat{e}_{3}$. With respect to pairing order parameters, its expectation value within the unprojected state
$\Psi_{0}$ can be written as
\begin{widetext}
\begin{flalign}\label{eq:paringw.r.t.order}
\langle\overrightarrow{\Delta}_{\mathbf{R}_{i}+\vec{\tau}_{1},\mathbf{R}_{i}}\rangle_{0}=\left(i(\tilde{\Delta}_{12}-\tilde{\Delta}_{34}),(\tilde{\Delta}_{12}+\tilde{\Delta}_{34}),-i(\tilde{\Delta}_{14}-\tilde{\Delta}_{23}),i(\tilde{\Delta}_{13}-\tilde{\Delta}_{24}),(\tilde{\Delta}_{13}+\tilde{\Delta}_{24}),-i(\tilde{\Delta}_{14}+\tilde{\Delta}_{23})\right).
\end{flalign}
\end{widetext}
Here $\mathbf{R}_{i}$ represents the site of A sublattice of the honeycomb lattice. We insert the flavor configuration
$\tilde{\Delta}=\tilde{\Delta}_{12}=\tilde{\Delta}_{34}=\tilde{\Delta}_{14}=\tilde{\Delta}_{23}=\tilde{\Delta}_{13}=-\tilde{\Delta}_{24}$ introduced in our paper into Eq.
(\ref{eq:paringw.r.t.order}) to obtain
\begin{flalign}\label{eq:paringw.r.t.orderD1}
\langle\overrightarrow{\Delta}_{\mathbf{R}_{i}+\vec{\tau}_{1},\mathbf{R}_{i}}\rangle_{0}=&\left(0,2\tilde{\Delta},0,2i\tilde{\Delta},0,-2i\tilde{\Delta}\right)
                        \nonumber\\
=&\langle\tilde{c}_{i}^{t}\left(0, i \sigma^{02}, 0, -\sigma^{23}, 0, \sigma^{21}\right) \tilde{c}_{j}\rangle_{0}.
\end{flalign}
Using $v_{1}$ and $v_{2}$ as the orbital index, we choose the representation $\tilde{c}_{j}=(c_{j\uparrow v_{1}},c_{j\uparrow v_{2}},c_{j\downarrow v_{1}},c_{j\downarrow v_{2}})^{t}$.
According to Eq. (\ref{eq:pairing field relation}), the previous three components of Eq. (\ref{eq:paringw.r.t.orderD1}) describe spin triplet with $d_{y}\ne0$, $d_{x}=d_{z}=0$ and
orbital singlet, where $\mathbf{d}$ is the vector defined by triplet pairing
\begin{flalign}
\hat{\Delta}(\mathbf{ k})=&i(\mathbf{ d}(\mathbf{ k})\cdot\hat{\sigma})\sigma^{2}
                      \nonumber\\
=&\left(\begin{array}{cc}-d_{x}(\mathbf{ k})+id_{y}(\mathbf{ k})&d_{z}(\mathbf{ k})\\d_{z}(\mathbf{ k})&d_{x}(\mathbf{ k})+id_{y}(\mathbf{ k})
\end{array}\right).
\label{eq:tripletpairing}
\end{flalign}
The last three components of Eq. (\ref{eq:paringw.r.t.orderD1}) describe spin singlet and orbital triplet with $d_{y}=0$, $d_{x}\ne0$, $d_{z}\ne0$. Note that the spin singlet pairing and
the spin triplet pairing coexist simultaneously in this flavor configuration. If we insert another flavor configuration
$\tilde{\Delta}=\tilde{\Delta}_{12}=-\tilde{\Delta}_{34}=\tilde{\Delta}_{14}=-\tilde{\Delta}_{23}=\tilde{\Delta}_{13}=\tilde{\Delta}_{24}$ into Eq. (\ref{eq:paringw.r.t.order}), we will
obtain
\begin{flalign}\label{eq:paringw.r.t.orderD2}
\langle\overrightarrow{\Delta}_{\mathbf{R}_{i}+\vec{\tau}_{1},\mathbf{R}_{i}}\rangle_{0}=&\left(2i\tilde{\Delta},0,-2i\tilde{\Delta},0,2\tilde{\Delta},0\right)
                    \nonumber\\
=&\langle\tilde{c}_{i}^{t}\left(-\sigma^{32}, 0, \sigma^{12}, 0, i\sigma^{20}, 0\right) \tilde{c}_{j}\rangle_{0}.
\end{flalign}
This indicates that there is symmetry between the spin index and the orbital index. As mentioned in our paper, there is the gauge redundancy. Namely, $c_{1}\to
c_{1}^{\prime}=c_{1}\mathrm{e}^{i(\theta_{1}+\theta_{2}+2\theta_{4})/2}$, $c_{2}\to c_{2}^{\prime}=c_{2}\mathrm{e}^{i(\theta_{1}-\theta_{3}+2\theta_{4})/2}$, $c_{3}\to
c_{3}^{\prime}=c_{3}\mathrm{e}^{i(\theta_{2}-\theta_{3}+2\theta_{4})/2}$, and $c_{4}\to c_{4}^{\prime}=c_{4}\mathrm{e}^{i\theta_{4}}$, then we obtain
$\tilde{\Delta}_{12}=\mathrm{e}^{-i\theta_{1}}\tilde{\Delta}_{34}$, $\tilde{\Delta}_{14}=\mathrm{e}^{-i\theta_{3}}\tilde{\Delta}_{23}$, and
$\tilde{\Delta}_{13}=-\mathrm{e}^{-i\theta_{2}}\tilde{\Delta}_{24}$. And insert this general flavor configuration into Eq. (\ref{eq:paringw.r.t.order}) to obtain
\begin{widetext}
\begin{flalign}\label{eq:paringw.r.t.orderarb}
\langle\overrightarrow{\Delta}_{\mathbf{R}_{i}+\vec{\tau}_{1},\mathbf{R}_{i}}\rangle_{0}=\left(i\tilde{\Delta}_{12}(1-\mathrm{e}^{i\theta_{1}}),\tilde{\Delta}_{12}(1+\mathrm{e}^{i\theta_{1}}),-i\tilde{\Delta}_{14}(1-\mathrm{e}^{i\theta_{3}}),i\tilde{\Delta}_{13}(1+\mathrm{e}^{i\theta_{2}}),\tilde{\Delta}_{13}(1-\mathrm{e}^{\theta_{2}}),-i\tilde{\Delta}_{14}(1+\mathrm{e}^{i\theta_{3}})\right).
\end{flalign}
\end{widetext}
We can clearly see that when $\theta_{i}\ne0,\pi$, ($i=1,2,3$), the six components of $\langle\overrightarrow{\Delta}_{\mathbf{R}_{i}+\vec{\tau}_{1},\mathbf{R}_{i}}\rangle_{0}$ are
simultaneously nonvanishing.


\end{document}